\documentclass[fleqn,twoside]{article}
\usepackage{espcrc2}
\usepackage{epsfig}

\newcommand{\AmS}{{\protect\the\textfont2
  A\kern-.1667em\lower.5ex\hbox{M}\kern-.125emS}}

\newcommand{\beq}{\begin{equation}}
\newcommand{\eeq}{\end{equation}}
\newcommand{\bea}{\vspace{0.25cm}\begin{eqnarray}}
\newcommand{\eea}{\end{eqnarray}}

\newcommand{\rr}{\mbox{{\boldmath
$\rho$}}}
\newcommand{\ta}{\mbox{{\boldmath
$\tau$}}}

\newcommand{\qb}{\mbox{{\bf
q}}}

\newcommand{\rb}{\mbox{{\bf
r}}}

\def\lsim{\mathrel{\rlap{\lower4pt\hbox{\hskip1pt$\sim$}}
    \raise1pt\hbox{$<$}}}         
\def\gsim{\mathrel{\rlap{\lower4pt\hbox{\hskip1pt$\sim$}}
    \raise1pt\hbox{$>$}}}         

\title{
Light-cone path integral approach to the induced
radiation in QED and QCD: basic concepts and recent applications
}

\author{B.G. Zakharov \address[Landau]{
 L.D.~Landau Institute for Theoretical Physics,
        GSP-1, 117940,\\ Kosygina Str. 2, 117334 Moscow, Russia}}
       
\begin{document}

\begin{abstract}
I discuss the basic ideas of the light-cone path integral approach
to the induced radiation in QED and QCD and recent applications
to the induced parton energy loss.
\vspace{1pc}
\end{abstract}

\maketitle

\noindent {\bf 1}.
The induced radiative energy loss 
and the Landau-Pomeranchuk-Migdal (LPM)
effect \cite{LP,Migdal} in QED and QCD have attracted much
attention in recent years, see \cite{Klein,BSZ,KW}
and references therein.
It is mainly because of the first accurate measurements of the 
LPM effect in QED at SLAC \cite{SLAC}, and the 
possibility to use jet quenching for probbing
the quark-gluon plasma (QGP) produced in high-energy $AA$-collisions
\cite{BSZ,GLV1,BDMS_quenching,Z2}.
The most general approach to the induced radiation applicable in both QED
and QCD  is the light-cone path integral (LCPI) approach
\cite{Z1} (see also \cite{Z_YAF,Z3,BSZ}). 
It accurately treats  the mass and finite-size effects, and applies
at arbitrary strength of the LPM effect. This approach gives excellent
description of the SLAC \cite{SLAC} and SPS \cite{SPS} data on the 
photon radiation from 
high-energy electrons \cite{Z_SLACSPS}.
In this talk I discuss the basic concepts and recent applications of the
LCPI approach.

\noindent {\bf 2}.
The starting point of the LCPI formalism is the representation of the
wave functions of high-energy free particles
in the form (it is assumed that the angles between the $z$ axis and velocities 
are small)
\beq
\psi_{j}(\rb)=\exp(iE_{j}z)\hat{U}_{j}\phi_{j}(\rr,z)\,,
\label{eq:10}
\eeq
where $\rr=\rb_{\perp}$, $\hat{U}_{j}$ is a spin operator, 
and $\phi_{j}$ satisfies the 
Schr\"odinger equation
\beq
i\frac{\partial \phi_{j}(\rr,z)}{\partial z}=
\left[-\frac{1}{2\mu_{j}}
\left(\frac{\partial}{\partial \rr}\right)^{2}+
\frac{m_{j}^{2}}{2\mu_{j}}\right]\phi_{j}(\rr,z)\,.
\label{eq:20}
\eeq
The Schr\"odinger mass $\mu_{j}$ in (\ref{eq:20})
equals the particle energy $E_{j}$. 
The $z$-evalution of $\phi_{j}$ is described by the Green's 
function for the Schr\"odinger equation (\ref{eq:20}), $K_{j}$.
In vacuum the amplitude squared $|\langle bc|T|a\rangle|^{2}$ for the
$a\rightarrow bc$ transition is given by the diagram shown in Fig.~1a,
where the $\rightarrow$ ($\leftarrow$) lines correspond to 
$K_{j}$ ($K_{j}^{*}$), 
the dashed lines show the
initial (at $z_{i}\sim -\infty$) and final (at $z_{f}\sim \infty$) 
transverse density matrices, $\rho_{i}$, $\rho_{f}$, and  
the integration over the transverse coordinates of the endpoints 
(and vertices) and the longitudinal coordinates of the vertices 
(below $z_{1}$ and $z_{2}$ for upper and lower
parts) is implicit.
In vacuum $\langle bc|T|a\rangle=0$ (if the particle $a$ is not produced
in a hard reaction, and  
$m_{a}\le m_{b}+m_{c}$), however, in an external potential
it is not the case.  
\begin{figure}
\vspace*{-4.6cm}
\hspace*{-1.5cm}
\epsfig{file=F1.epsi,height=15.5cm}
\vspace*{-10.3cm}
\caption{}\vspace*{-1cm}
\end{figure}

To evaluate the induced $a\rightarrow bc$ transition in an amorphous
medium one should sum $t$-channel exchanges between the fast particles 
and particles in the medium and perform averaging over
the medium states. The key idea of the LCPI approach is to represent
all $K_{j}$ in the Feynman path integral form, and perform
averaging over the potential at the level of the integrands before integrating 
over the trajectories. 
After averaging over the medium states the interaction of 
the fast particles with the medium 
turns out to be translated to an effective interaction between trajectories.
Both in QED and QCD (note that in QCD we work at the level of two-gluon 
$t$-channel exchanges) 
from the viewpoint of this interaction the trajectories 
corresponding
to $K_{j}^{*}$ may be viewd as antiparticle trajectories (we call 
$\rightarrow$ and $\leftarrow$ lines  ``particle'' and
``antiparticle'' trajectories). 
In medium the effective Lagrangian for the path integral form of the 
diagram 1a reads
\beq
L_{eff}
=L_{0}^{p}(\dot{\ta}_{p})-L_{0}^{\bar{p}}(\dot{\ta}_{\bar{p}})+
L_{int}(\ta_{p},\ta_{\bar{p}}),
\label{eq:30}
\eeq
where $\ta_{p}$, $\ta_{\bar{p}}$ are the sets of the 
transverse coordinates for ``particles'' and ``antiparticles'', 
$L_{0}^{p}$ and $L_{0}^{\bar{p}}$ are the corresponding free Lagrangians,
say,  
$L_{0}^{p}(\dot{\ta}_{p})=\sum_i\mu_{i}\dot{\rr}_{i}^{2}/2$. The interaction
term reads
$L_{int}(\ta_{p},\ta_{\bar{p}})=
{in(z)\sigma_{X}(\ta_{p},\ta_{\bar{p}})}/{2}$, where
$\sigma_{X}$ is the diffractive operator for $X=$ 
``particles''+''antiparticles'' system scattering off 
a particle in medium, $n(z)$ is the number density of the medium.
In QED $\sigma_{X}$ is simply the cross section, but
in QCD for the 4-body part ($z_{2}<z<z_{f}$) $\sigma_{X}$ 
acts as an operator in color space.
The one-particle spectrum, say, integrated over $\qb_{c}$ (hereafter 
$\qb$ stands for the transverse momentum), may be represented
in a form which does not contain the 4-body part at all. Since 
integration over $\qb_{c}$ gives $\delta(\rr_{c}-\rr_{\bar{c}})$, the diagram 
1a
is transformed into 1b in this case.
The spectrum correponding to the diagram 1b reads
\bea
\frac{dP}
{dxd\qb_{b}}\!\!&\!\!=\!\!&\!\!2\mbox{Re}\int_{z_{i}}^{z_{1}}\!\!\!
dz_{1}
\int_{z_{1}}^{z_{f}}\!\!\!
dz_{2}\nonumber\\ &\times\!\!&\!\hat{g}
\langle \rho_{f}|
\hat{S}_{b\bar{b}}\otimes
\hat{S}_{bc\bar{a}}
\otimes
\hat{S}_{a\bar{a}}
|\rho_{i}\rangle,
\label{eq:40}
\eea
where $x=x_{b}=p_{b,z}/p_{a,z}$, $\hat{g}$ is the vertex factor, 
$\hat{S}_{a\bar{a}}$, $\hat{S}_{b\bar{b}}$, $\hat{S}_{bc\bar{c}}$
are the evolution operators for the corresponding $L_{eff}$.
For the two-body parts the path integrals can be taken analytically, say, 
\vspace*{-.1cm}
\bea
\lefteqn{
\langle \rr_{a}',\rr_{\bar{a}}',z'|\hat{S}_{a\bar{a}}|
\rr_{a},\rr_{\bar{a}},z
\rangle
=K_{a}(\rr_{a}',z'|\rr_{a},z)}
\nonumber\\ \times
\lefteqn{
K_{\bar{a}}^{*}
(\rr_{\bar{a}}',z'|\rr_{\bar{a}},z)
\Phi_{a\bar{a}}(\rr_{a}'-\rr_{\bar{a}}',z'|\rr_{a}-\rr_{\bar{a}},z),}
\label{eq:50}
\eea
where
$\Phi_{a\bar{a}}\!\!=\!\!\exp{[-\frac{1}{2}\int_{z}^{z'}\!\!\! dz n(z)
\sigma_{a\bar{a}}(\rr_{a}(z)-\rr_{\bar{a}}(z))]}$ 
should be evaluated for the straight trajectories of $a$ and $\bar{a}$.
For the $bc\bar{a}$-part one can write 
$\int{D}\rr_{b}{D}\rr_{c}{D}\rr_{\bar{a}}$
as $\int{D}\rr{D}\rr_{a}{D}\rr_{\bar{a}}$,
where $\rr=\rr_{b}-\rr_{c}$, and
$\rr_{a}=x_{b}\rr_{b}+x_{c}\rr_{c}$ is the center-of-mass coordinate of the 
$bc$ system. The   
$\rr_{a}$ and $\rr_{\bar{a}}$ integrals can be taken analytically, and
in the new variables $S_{bc\bar{a}}$ can be represented
as
\bea
\lefteqn{
\langle \rr_{a}',\rr_{\bar{a}}',\rr',z'|\hat{S}|\rr_{a},\rr_{\bar{a}},
\rr,z\rangle
=K_{a}(\rr_{a}',z'|\rr_{a},z)}\nonumber\\ \lefteqn{\times
K_{\bar{a}}^{*}
(\rr_{\bar{a}}',z'|\rr_{\bar{a}},z)
{\cal{K}}(\rr',z'|\rr,z),}
\label{eq:60}
\eea
where ${\cal{K}}$ is the Green's function for the Hamiltonian
\beq
\hat{H}=
-\frac{1}{2M(x)}\,
\left(\frac{\partial}{\partial \rr}\right)^{2}
+v(z,\rr) +\frac{1}{L_{f}}\,.
\label{eq:70}
\eeq
Here
$
v(z,\rr)=-i{n(z)\sigma_{bc\bar{a}}(\rr,\rr_{a}-\rr_{\bar{a}})}/{2}
$
should be evaluated for the straight trajectories $\rr_{a}$, $\rr_{\bar{a}}$,
$M(x)=E_{a}x(1-x)$,
$L_{f}={2M(x)}/\epsilon^{2}$ is the formation length,
$\epsilon^{2}={[m_{b}^{2}x_{c}+m_{c}^{2}x_{b}-m_{a}^{2}x_{b}x_{c}]}$.
Having (\ref{eq:50}) and (\ref{eq:60}), one can obtain
\bea
\lefteqn{
\frac{dP}{dx d\qb_{b}}\!=\! 
\frac{2}{(2\pi)^2} {\rm Re}\!\! \int\! d \ta\!  
\exp (-i \qb_b \cdot \ta)\!\!
 \int^{z_f}_{z_i}\!\! dz_1\!\!}
\nonumber\\
\lefteqn{\times
\int^{z_f}_{z_1}\!\! 
dz_2 
\hat{g}  \Phi_f (\ta , z_2){\cal{K}}(\ta, z_2 | 0 , z_1) \Phi_i 
(x\ta , z_1),
}
\label{eq:80}
\eea
\vspace*{-.2cm}
$$
\Phi_i ( \ta , z_1 )\!\!=\!\!\exp \left.\Big[ - \frac{\sigma_{a \bar a} 
(\ta)}{2}\,
\int^{z_1}_{z_i}\!\! dz  n (z) \right.\Big],
$$
$$
\Phi_f (\ta , z_2)\!\! =\!\! \exp \left.\Big[ - 
\frac{\sigma_{b \bar b} (\ta)}{2} \, \int^{z_f}_{z_2}\!\!
 dz n(z)\right.\Big].
$$
The contribution stemming from the region of large $|z_{1,2}|$ 
in (\ref{eq:80}) can be 
expressed via the light-cone wave-function of the  $a\rightarrow bc$
transition in vacuum, $\Psi^{bc}_a$, \cite{BSZ}.
Then (\ref{eq:80}) takes the form
$$
\frac{dP}{dxd\qb_{b}}\!=\!  \frac{2}{(2\pi)^2} {\rm Re}\!\!  \int\!\!  d\ta  
\exp (- i\qb_{b}\ta) 
$$\vspace*{-3mm}
$$
\times\int^{z_f}_{z_i}\!\!  dz_1\!  \int^{z_f}_{z_1}\!\! dz_2
\hat{g}\{ \Phi_f(\ta , z_2) [{\cal{ K}} (\ta , z_2 | 0 , z_1)
$$\vspace*{-3mm}
$$
- {\cal{K}}_{v} (\ta , z_2 | 
0 , z_1)] \Phi_i (x\ta , z_1 )
$$\vspace*{-3.5mm}
$$
 +
 [ \Phi_f (\ta , z_2 ) - 1 ]{\cal{ K}}_{v} (\ta , z_2 | 0 , z_1) 
[\Phi_i (x\ta , z_1 ) - 1 ]\}
$$
\vspace*{-3.5mm}
$$
 + \frac{1}{(2\pi)^{2}}\int d\ta d\ta'
\exp (- i\qb_b  \ta)
\Psi^{bc*}_a (x , \ta'-\ta)
$$
\vspace*{-5mm}
\beq
\times
 \Psi^{bc}_a (x , \ta')
\left[ \Phi_f (\ta , z_i ) + \Phi_i (x\ta , z_f ) - 2 \right] ,
\label{eq:90}
\eeq
where ${\cal{K}}_{v}$ is the Green's function for $v=0$.
For $L_{f}\gg L$ ($L$ is the thickness of the medium) 
using the representation of $\Psi_{a}^{bc}$ via
the $z$-integral of ${\cal{K}}$ esteblished in \cite{Z_SLAC1} one can 
represent (\ref{eq:90}) as
\bea
\frac{dP}{dxd\qb_{b}}\!=\!  \frac{1}{(2\pi)^2} {\rm Re}\!\!  
\int d\ta d\ta'
\exp (- i\qb_b  \ta)\nonumber\\ \times
\Psi^{bc*}_{a} (x , \ta'-\ta)
 \Psi^{bc}_{a} (x , \ta')\nonumber\\ \times
\left[2\Gamma_{bc\bar{a}}(\ta',x\ta)\!- \!\Gamma_{b\bar{b}}(\ta)\!-\!
\Gamma_{a\bar{a}} (x\ta )\right] ,
\label{eq:100}
\eea
where $\Gamma_{h}=
1-\exp{[-\frac{\sigma_{h}}{2}\int_{-\infty}^{\infty}dz n(z)]}$.

From (\ref{eq:90}) one obtains for the $x$-spectrum \cite{Z1}
\bea
\frac{dP}{dx}\! =\! 2 {\rm Re} \!\int^{z_f}_{z_i}\! dz_1\! 
\int^{z_f}_{z_1}\!
dz_2 \hat{g} \left[{\cal{K}} (\rr_2 , z_2 | \rr_1 , z_1)\right.
\nonumber\\
 -\left. {\cal{K}}_{v} 
(\rr_2 , z_2 | \rr_1 , z_1 )\right]{\Big|}_{\rr_1 = \rr_2 = 0} .
\label{eq:110} 
\eea
In \cite{Z_SLAC1}, separating the $N=1$ rescattering, 
we have represented (\ref{eq:110})
as a sum of the Bethe-Heitler spectrum plus an absorptive correction
responsible for the LPM effect.
This form has been used \cite{Z_SLACSPS} for successful description of 
the SLAC \cite{SLAC} and SPS \cite{SPS} data
on the LPM effect in photon bremsstrahlung from high energy electrons.

For a particle produced in the medium it is convenient to 
rewrite (\ref{eq:110}) in another form. For gluon emission from a quark
this new form reads \cite{Z_jq}
\beq
\frac{d P}{d
x}=
\int_{0}^{L}\! d z\,
n(z)
\frac{d
\sigma_{eff}^{BH}(x,z)}{dx}\,,
\label{eq:120}
\eeq
\vspace*{-.5cm}
\bea
\lefteqn{
\frac{d
\sigma_{eff}^{BH}(x,z)}{dx}=\mbox{Re}\!\!
\int_{0}^{z} dz_{1}\!\!\int_{z}^{\infty}\!\!dz_{2}\!\!\int 
\!\!d\rr
\hat{g}}
\nonumber\\ 
\lefteqn{
\times
{\cal{K}}_{v}(\rr_{2},z_{2}|\rr,z)
\sigma_{3}(\rho)
{\cal{K}}(\rr,z|\rr_{1},z_{1}){\Big|}_{\rr_{1}\!=\!\rr_{2}\!=\!0}\,,}
\label{eq:130}
\eea
where $\sigma_{3}=\sigma_{gq\bar{q}}$,
$L$ is the quark pathlength in the medium.
The $d\sigma^{BH}_{eff}/dx$ (\ref{eq:130}) can be viewed as an 
effective Bethe-Heitler
cross section which accounts for the LPM and finite-size effects.
Neglecting spin-flip transition (\ref{eq:130}) can be written as
\bea
\lefteqn{
\frac{d
\sigma_{eff}^{BH}(x,z)}{dx}=-\frac{\alpha_{s}P_{q}^{gq}(x)}{\pi M(x)}}
\nonumber\\
\lefteqn{\phantom{+++} 
\times\mbox{Im}
\int_{0}^{z} d\xi
\left.\frac{\partial }{\partial \rho}
\left(\frac{F(\xi,\rho)}{\sqrt{\rho}}\right)
\right|_{\rho=0}\,,}
\label{eq:140}
\eea
where 
$P(x)_{q}^{gq}=C_{F}[1+(1-x)^{2}]/x$ is the usual splitting function, and
$F$ is the solution to the radial Schr\"odinger 
equation for the azimuthal quantum number $m=1$
\bea
\lefteqn{
i\frac{\partial F(\xi,\rho)}{\partial \xi}=
\left[-\frac{1}{2M(x)}\left(\frac{\partial}{\partial \rho}\right)^{2}
\right.}
\nonumber\\
\lefteqn{
\left.
-i\frac{n(z-\xi)\sigma_{3}(\rho)}{2}+
\frac{4m^{2}-1}{8M(x)\rho^{2}}
+\frac{1}{L_{f}}
\right]F(\xi,\rho)\,.}
\label{eq:150}
\eea
The boundary condition for $F(\xi,\rho)$ reads
$F(\xi=0,\rho)=\sqrt{\rho}\sigma_{3}(\rho)
\epsilon K_{1}(\epsilon \rho)$, where 
$K_{1}$ is the Bessel function.

\begin{figure}
\vspace*{-2cm}
\epsfig{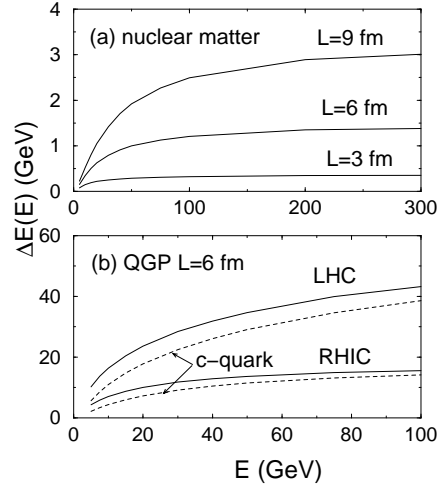}
\vspace*{-1.cm}
\caption{The quark energy loss in the nuclear matter (a) for 
$L=3$, 6, 9 fm 
and expanding QGP (b) for $L=6$ fm for RHIC and LHC.
}
\end{figure}
\noindent {\bf 3}. The formulas (\ref{eq:120}), (\ref{eq:140}) are 
convenient for numerical calculations with an accurate parametrization
of $\sigma_{3}$. Note that the widely used oscillator approximation
$\sigma_{3}(\rho)\propto \rho^{2}$ is too crude for the induced gluon
emission \cite{Z_OA,Z_jq}. We perform calculations for running
$\alpha_{s}$ frozen at $\alpha_{s}=0.7$ at low momenta 
(for incorporation of running $\alpha_{s}$ see \cite{Z_jq}). 
In Fig.~2 we show the energy dependence 
of the quark energy loss $\Delta E=E\int dxxdP/dx$ for the cold nuclear 
matter 
and for expanding QGP for RHIC and LHC conditions.
For nuclear matter we take $m_{g}=0.75$ GeV obtained from the 
analysis of the low-$x$ proton structure function within the dipole 
BFKL equation \cite{NZZ,NZ_HERA}.
It agrees well with the inverse gluon correlation radius in the QCD vacuum
\cite{Shuryak1}.   
For QGP we use $m_{g}=0.4$ GeV obtained in 
the quasiparticle model from
the lattice data in \cite{LH}. 
For the light quark we take $m_{q}=0.3$ GeV. For QGP
this quark mass was obtained in \cite{LH}. Note that 
our results are not very sensitive to the light 
quark mass. For the QGP we also show the results for 
$c$-quark, $m_{q}=1.5$ GeV.
The Debye screening mass, $\mu_{D}$, in QGP was obtained
from the perturbative relation $\mu_{D}=\sqrt{2}m_{g}\approx 0.57$ GeV.
We use the Bjorken 
model for the QGP expansion with
$T^{3}\tau=T_{0}^{3}\tau_{0}$
and the initial conditions suggested 
in \cite{FMS}:
$T_{0}=446$ MeV and $\tau_{0}=0.147$ fm for RHIC (for $Au+Au$ at
$\sqrt{s}=200$ GeV), and
$T_{0}=897$ MeV and 
$\tau_{0}=0.073$ fm for LHC ($Pb+Pb$ at $\sqrt{s}=5.5$ TeV).
\begin{figure}
\vspace*{-3.7cm}
\epsfig{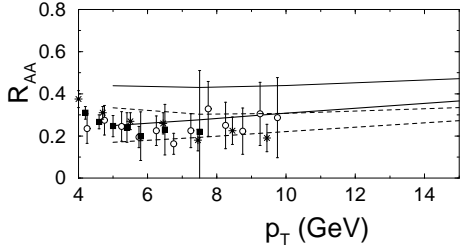}
\vspace*{-3.7cm}
\caption[.]{
The nuclear modification factor 
for central $Au+Au$ collisions at $\sqrt{s}=200$ GeV
for quark (solid line) and 
gluon (dashed line)
jets obtained with $m_{g}=0.4$ GeV (thick lines) and $m_{g}=0.75$ GeV 
(thin lines).
The experimental points (from \cite{RHIC_data}) are for:
circle - $Au+Au\rightarrow \pi^{0}+X$ (0-10\% central) 
[PHENIX Collaboration], 
square - $Au+Au\rightarrow h^{\pm}+X$ (0-10\% central)
 [PHENIX Collaboration],
star - $Au+Au\rightarrow h^{\pm}+X$ (0-5\% central)
[STAR Collaboration].
}
\end{figure}
For the QGP we take $L=6$ fm. It is a about 
the life-time of the QGP (and mixed) phase $\tau_{max}\sim R_{A}$
for central heavy-ion collisions, since  due to the transverse expansion 
the hot QCD matter should cool quickly at $\tau\gsim R_{A}$.

$\Delta E$ for nuclear matter obtained using (\ref{eq:120})-(\ref{eq:150})
with an accurate parametrization of the imaginary potential by a
factor of about 2 smaller than that obtained previously
\cite{Z2} in the  
oscillator approximation.
Our $\Delta E$ for nuclear matter is considerably smaller
than prediction of \cite{Wang1} in the HERMES \cite{HERMES} energy region 
$E\lsim 20$ GeV. The small $\Delta E$ shows that for the HERMES conditions 
the effects
of hadron absorption and string tension should be more important
than the gluon emission since the hadron formation time
$l_{f}$ is about the nuclear size. Note that the estimate 
$l_{f}\sim 40$ fm for $E=8$ GeV given in \cite{Wang1} is 
absolutely unrealistic. The authors of \cite{Wang1} do not pay any attention to the
fact that a quark with such an energy should be stopped at
$L\lsim 8$ fm since the string tension is $\sim 1$ GeV/fm.

One can see from Fig.~2b that for LHC conditions the jets with 
$E\lsim 20$ GeV should practically be absorbed in the QGP. 
It means that the surface jet production dominates.
The energy loss
for $c$-quark is smaller than that for light quarks by a factor of $\sim 2$
at $E\sim 5-10$ GeV.

The effect of parton energy loss on the high-$p_{T}$ hadron
production in $A+A$ collisions can approximately be described in 
terms of effective
hard partonic cross sections which account for the induced gluon emission
\cite{BDMS_quenching}.
Using the power-low parametrization for cross section of quark 
production in $p+p$ collisions $\propto 1/p_{T}^{n(p_{T})}$
the nuclear modification factor
\beq
R_{AA}(p_{T})=
\frac{d\sigma^{AA}(p_{T})/dydp_{T}^{2}}
{N_{bin}d\sigma^{pp}(p_{T})/dydp_{T}^{2}}
\label{eq:160}
\eeq
can be written as \cite{Z_jq}
\bea
\lefteqn{
R_{AA}(p_{T})\!\approx\! P_{0}(p_{T})\!\!+\!\!\frac{1}{J(p_{T})}\!
\int_{0}^{1}\!\!\! dz z^{n(p_{T})-2}\!D_{q}^{h}(z,\frac{p_{T}}{z})}\nonumber\\
\lefteqn{\times
\!\!
\int_{0}^{1}\!\! dx(1-x)^{n(p_{T}/{z})-2} 
\frac{dI(x,\frac{p_{T}}{z(1-x)})}{dx},}
\label{eq:170}
\eea
where
$
J(p_{T})=\int_{0}^{1}dz z^{n(p_{T})-2} D_{q}^{h}(z,
\frac{p_{T}}{z})
$, $P_{0}$ is the probability of quark propagation without induced 
gluon emission,
$dI(x,p_{T})/dx$ is the probability distribution in 
the quark energy loss for a quark with $E=p_{T}$,
$D_{q}^{h}(z,p_{T}/z)$ is the quark fragmentation function.
A convenient parametrization for $P_{0}$ and $dI(x,p_{T})/dx$ in terms of 
$dP/dx$
(similar to that used for photon emission in \cite{Z_SLACSPS})
is given in \cite{Z_jq}.
\begin{figure}
\vspace*{-3.7cm}
\epsfig{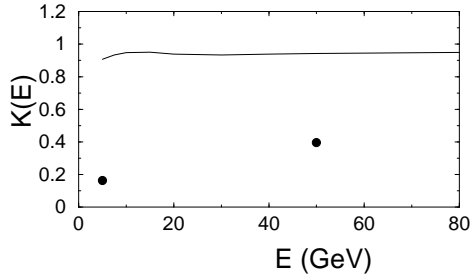}
\vspace*{-3.7cm}
\caption[.]{The kinematic $K$-factor for $N=1$ rescattering for the QGP,
$L=5$ fm, $n(z)=const$. Solid line shows our results.  
The points show the GLV \cite{GLV1} predictions.
}
\end{figure}
In Fig.~3 we compare the theoretical $R_{AA}$ 
with the RHIC data on 
$Au+Au$ collisions at $\sqrt{s}=200$ GeV \cite{RHIC_data}.
The theoretical curves have 
been obtained for $L=4.9$ fm. It is the typical parton pathlength in
the QGP (and mixed) phase for $\tau_{max}=6$ fm. 
The results for the quark 
and gluon jets are shown separately
(for $\sqrt{s}=200$ GeV the quark and gluon contributions 
are comparable). 
The suppression is somewhat stronger for gluon jets.
As for Fig.~2b the calculations are performed for $m_{q}=0.3$ GeV
and $m_{g}=0.4$ GeV. To illustrate the $m_{g}$-dependence 
we also show the results for $m_{g}=0.75$ GeV
(with the same $\mu_{D}$ as for $m_{g}=0.4$ GeV), which
seems to be reasonable for the mixed phase and 
for gluons with large formation length $L_{f}\gsim L$. 
The above two values of $m_{g}$ give reasonable 
lower and upper limits of the infrared cutoff for the induced gluon
emission for RHIC conditions. 
The theoretical $R_{AA}$ for 
$m_{g}=0.4$ GeV is in quite good agreement with the experimental one.

\noindent{\bf 4}.
The LCPI approach neglects the kinematic bounds.
The GLV group \cite{GLV1} reported that the finite kinematic limits
suppress strongly the energy loss even at $E\sim 500$ GeV.
To check it in \cite{Z_kinb} we have calculated the dominating $N=1$ 
rescattering contribution to the quark energy 
loss using the ordinary diagram language. We have
obtained a small kinematic effect. To demonstrate it in Fig.~4 
we show the kinematic $K$-factor $K=\Delta E_{f.l.}/\Delta E_{i.l.}$, 
where $\Delta E_{f.l.}$ and $\Delta E_{i.l.}$ are the energy losses for 
finite and infinite kinematic limits.
Our calculations are performed for the kinematic limits as in \cite{GLV1},
also similarly to \cite{GLV1} we use fixed $\alpha_{s}$, $n(z)=const$, 
$L=5$ fm, and $m_{g}=\mu_{D}=0.5$ GeV. 
One sees that our $K$-factor contrary to the GLV predictions is 
close to unity even
for $E\sim 5$ GeV. It says that the LCPI approach has a quite good accuracy for
$E\gsim 5$ GeV. 

{\small
I thank the Organizers of Diffraction'04 for the hospitality
and financial support during the Workshop.}

\end{document}